\documentclass[aps,floatfix,nofootinbib,showpacs,twocolumn,superscriptaddress]{revtex4}
\usepackage{amsmath}
\usepackage{amssymb}
\usepackage{amsthm}
\usepackage{dcolumn}
\usepackage{epsfig}
\usepackage{graphics}
\usepackage{graphicx}
\usepackage{amsmath}
\usepackage{longtable}
\usepackage{color}

\definecolor{darkgreen}{rgb}{0,0.5,0}
\definecolor{purple}{rgb}{0.5,0,0.5}
\definecolor{nblue}{rgb}{0.0,0.0,0.50}
\definecolor{scarlet}{rgb}{1.0,0.2,0}
\usepackage[colorlinks=true, pdfstartview=FitV, linkcolor=purple, citecolor= purple, urlcolor=blue]{hyperref}

\begin{document}

%\title{Quiddity of the chiral condensate}
\title{Essence of the vacuum quark condensate}
%PhysRev always removes a leading article "the" or "an", etc.

\author{Stanley~J.~Brodsky}
\affiliation{SLAC National Accelerator Laboratory,
Stanford University, Stanford, CA 94309}
\affiliation{Centre for Particle Physics Phenomenology: CP$^3$-Origins, University of Southern Denmark, Odense 5230 M, Denmark}

\author{Craig~D.~Roberts} \affiliation{Physics Division, Argonne National
Laboratory, Argonne, Illinois 60439, USA} \affiliation{Department of Physics,
Peking University, Beijing 100871, China}

\author{Robert~Shrock}
\affiliation{C.N. Yang Institute for Theoretical Physics,
Stony Brook University, Stony Brook, NY 11794}

\author{Peter~C.~Tandy} \affiliation{Center for Nuclear Research, Department of
Physics, Kent State University, Kent OH 44242, USA}

\begin{abstract}
We show that the chiral-limit vacuum quark condensate is qualitatively equivalent to the pseudoscalar meson leptonic decay constant in the sense that they are both obtained as the chiral-limit value of well-defined gauge-invariant hadron-to-vacuum transition amplitudes that possess a spectral representation in terms of the current-quark mass.  Thus, whereas it might sometimes be convenient to imagine otherwise, neither is essentially a constant mass-scale that fills all spacetime.  This means, in particular, that the quark condensate can be understood as a property of hadrons themselves, which is expressed, for example, in their Bethe-Salpeter or light-front wavefunctions.
\end{abstract}


\maketitle

%\date{\today}

Non-zero vacuum expectation values of local operators; i.e., condensates, are introduced as parameters in QCD sum rules, which are used to estimate essentially nonperturbative strong-interaction matrix elements.  They are also basic to current algebra analyses.  It is widely held that such quark and gluon condensates have a physical existence, which is independent of the hadrons that express QCD's asymptotically realizable degrees-of-freedom; namely, that these condensates are not merely mass-dimensioned parameters in a theoretical truncation scheme, but in fact describe measurable spacetime-independent configurations of QCD's elementary degrees-of-freedom in a hadron-less ground state.

We share the view that these condensates are fundamental dynamically-generated mass-scales in QCD.  However, we shall argue that their measurable impact is entirely expressed in the properties of QCD's asymptotically realizable states; namely hadrons.
In taking this position we have assumed confinement, from which follows quark-hadron duality and hence that all observable consequences of QCD can, in principle, be computed using a hadronic basis.  Here, the term ``hadron'' means any one of the states or resonances in the complete spectrum of color-singlet bound-states generated by the theory.

We focus herein on $\langle 0 | \bar q q | 0 \rangle$, where $|0\rangle$ is viewed as some hadron-less ground state of QCD.  This is the vacuum quark condensate.  Its non-zero value is usually held to signal dynamical chiral symmetry breaking (DCSB), a concept of critical importance in QCD, whose connection with the dressed-quark propagator was anticipated \cite{Lane:1974he,Politzer:1976tv,Pagels:1978ba,Casher:1979vw,Banks:1979yr} (see also references therein).  As reviewed elsewhere (most recently, e.g., Refs.\,\cite{Fischer:2006ub,Roberts:2007jh,Chang:2010jq}),
%explained again recently \cite{Chang:2010jq},
DCSB is a remarkably efficient mass-generating mechanism, the origin of constituent-quark masses and intimately connected with confinement.  It is also the basis for the successful application of chiral-effective field theories (see, e.g., Refs.\,\cite{Bijnens:2006zp,Bernard:2006gx} for contemporary perspectives).  On the face of it, this seems far more than can be understood simply in terms of a non-zero vacuum expectation value $\langle 0| \bar q q |0\rangle$.

The notion that non-zero vacuum condensates exist and possess a measurable reality has long been recognized as posing a conundrum for the light-front formulation of QCD.  This formulation follows from Dirac's front form of relativistic dynamics \cite{Dirac:1949cp}, and is widely and efficaciously employed in perturbative and nonperturbative QCD \cite{Lepage:1980fj,Brodsky:1997de-Brodsky:2008pg}.  In the light-front formulation, the ground-state is a structureless Fock space vacuum, in which case it would seem to follow that DCSB is impossible.  In response, it was argued by Casher and Susskind \cite{Casher:1974xd} that, in the light-front framework, DCSB must be a property of hadron wavefunctions, not of the vacuum.  This thesis has also been explored in a series of recent articles
\cite{Brodsky:2009zd,Brodsky:2008be,Brodsky:2008tk}.

A non-zero spacetime-independent QCD vacuum condensate also poses a critical dilemma for gravitational interactions because it would lead to a cosmological constant some 45 orders of magnitude larger than observation.  As noted elsewhere \cite{Brodsky:2009zd}, this conflict is avoided if strong interaction condensates are properties of rigorously well-defined wavefunctions of the hadrons, rather than the hadron-less ground state of QCD.

Given the importance of DCSB and the longstanding puzzles described above, we will focus our attention on the vacuum quark condensate.  The essential issues become particularly clear in the context of the Gell-Mann--Oakes--Renner relation \cite{GellMann:1968rz,Dashen:1969eg}, which is usually understood as the statement
\begin{equation}
\label{gmor}
f_\pi^2 \, m_\pi^2 = - 2(m^u_\zeta+m^d_\zeta) \,
\langle \bar q q \rangle_\zeta^0,
\end{equation}
wherein $m_\pi$ is the pion's mass; $f_\pi$ is its leptonic decay constant; $m^q_\zeta$, with $q=u,d$, is the current-quark mass at a renormalization scale $\zeta$; and $\langle \bar q q \rangle_\zeta^0$ is the chiral-limit vacuum quark condensate, with a precise definition of the chiral limit given below in Eqs.\,(\ref{cl1}), (\ref{cl2}).
In arriving at Eq.\,(\ref{gmor}) using standard methods, one makes truncations; namely, soft-pion techniques \cite{adlerdashen} have been used to relate an in-pion matrix element of the current-quark mass-term to the vacuum quark condensate.  (NB.\ For technical simplicity, we will only explicitly consider the $SU(2)$-flavor case of two light quarks, with $m_u=m=m_d$.  A discussion of $SU(3)$-flavor and the $\eta$-$\eta^\prime$ complex is qualitatively identical and readily accomplished following Ref.\,\cite{Bhagwat:2007ha}.)

It is instructive to consider Eq.\,(\ref{gmor}) in another light.  Chiral symmetry and the pattern by which it is broken in QCD are expressed through the axial-vector Ward-Takahashi identity,
\begin{eqnarray}
\nonumber
&&P_\mu \Gamma_{5\mu}(k;P;\zeta) +  2 i m_\zeta \,\Gamma_5(k;P;\zeta)\\
& = & S^{-1}(k_+;\zeta) i \gamma_5 +  i \gamma_5 S^{-1}(k_-;\zeta) \,.
\label{avwtim}
\end{eqnarray}
Here $\Gamma_{5\mu}(k;P;\zeta)$ is the axial-vector vertex; $\Gamma_5(k;P;\zeta)$ is the pseudoscalar vertex; $k_\pm = k\pm P/2$; and $S(\ell;\zeta)$ is the dressed-quark propagator, which has the general form\footnote{We are using a Euclidean metric, with $\{\gamma_\mu,\gamma_\nu\} = 2\delta_{\mu\nu}$; $\gamma_\mu^\dagger = \gamma_\mu$; $\gamma_5= \gamma_4\gamma_1\gamma_2\gamma_3$; $a \cdot b = \sum_{i=1}^4 a_i b_i$; and $P_\mu$ timelike $\Rightarrow$ $P^2<0$.}
\begin{equation}
S(\ell;\zeta) = 1/[i\gamma\cdot \ell A(\ell^2;\zeta) + B(\ell^2;\zeta)]\,.
\end{equation}

Making use of the fact that the ground-state pion is the lowest-mass pole in both vertices if, and only if, chiral symmetry is dynamically broken, one can derive the following identity \cite{Maris:1997hd}, which is exact in QCD:
\begin{equation}
\label{gmorgen}
f_\pi \, m_\pi^2 =  2 m_\zeta \rho_\pi(\zeta)\, ,
\end{equation}
where (omitting $\zeta$ unless necessary for clarity or emphasis)
\begin{eqnarray}
%\nonumber
\label{fpigen}
\lefteqn{i f_\pi P_\mu = \langle 0 | \bar q \gamma_5 \gamma_\mu q |\pi \rangle} \\
\nonumber
& = & Z_2(\zeta,\Lambda)\; {\rm tr}_{\rm CD}
\int^\Lambda \!\!\!\! \mbox{\footnotesize $\displaystyle\frac{d^4 q}{(2\pi)^4}$} i\gamma_5\gamma_\mu S(q_+) \Gamma_\pi(q;P) S(q_-)\,, %\label{fpigen}
\\
\nonumber
\lefteqn{i\rho_\pi = -\langle 0 | \bar q i\gamma_5 q |\pi \rangle} \\
& = & Z_4(\zeta,\Lambda)\; {\rm tr}_{\rm CD}
\int^\Lambda \!\!\!\! \mbox{\footnotesize $\displaystyle\frac{d^4 q}{(2\pi)^4}$} \gamma_5 S(q_+) \Gamma_\pi(q;P) S(q_-) \,.\label{rhogen}
\end{eqnarray}
Here $\int^\Lambda \!\! \mbox{\footnotesize $\displaystyle\frac{d^4 q}{(2\pi)^4}$}$ represents a Poincar\'e-invariant regularization of the integral, with $\Lambda$ the ultraviolet regularization mass-scale,\footnote{In connection with Eq.\,(\protect\ref{M0mg0}) below, we describe how confinement and dynamical mass generation regulate the infrared domain in Eqs.\,(\protect\ref{fpigen}), (\protect\ref{rhogen}), thus also ensuring the absence of infrared divergences.
%how dynamical mass generation ensures the absence of infrared divergences.
}
and $\Gamma_{\pi}(k;P) $ is the pion's Bethe-Salpeter amplitude; viz.,
\begin{eqnarray}
\nonumber
\lefteqn{\Gamma_{\pi}(k;P) = \gamma_5 \left[ i E_\pi(k;P) + \gamma\cdot P F_\pi(k;P) \right.}\\
&& \left. + \gamma\cdot k \, G_\pi(k;P) - \sigma_{\mu\nu} k_\mu P_\nu H_\pi(k;P) \right].
\label{genGpi}
\end{eqnarray}
The quark wavefunction and Lagrangian mass renormalization constants, $Z_{2,4}(\zeta,\Lambda)$, respectively, depend on the gauge parameter in precisely the manner needed to ensure that the right-hand sides of Eqs.\,(\ref{fpigen}), (\ref{rhogen}) are gauge-invariant.  Moreover, $Z_2(\zeta,\Lambda)$ ensures that the right-hand side of Eq.\,(\ref{fpigen}) is independent of both $\zeta$ and $\Lambda$, so that $f_\pi$ is truly an observable; and $Z_4(\zeta,\Lambda)$ ensures that $\rho_\pi(\zeta)$ is independent of $\Lambda$ and evolves with $\zeta$ in just the way necessary to guarantee that the product $m_\zeta \rho_\pi(\zeta)$ is
renormalization-point-independent.

We now can discuss the chiral limit, which is well-defined in QCD since it is an asymptotically free, confining theory.  Recall that
\begin{equation}
\label{cl1}
Z_2(\zeta,\Lambda) \, m^{\rm bm}(\Lambda)=Z_4(\zeta,\Lambda) \, m_\zeta\,,
\end{equation}
where $m^{\rm bm}(\Lambda)$ is the Lagrangian bare-mass parameter.  Then, the chiral limit is defined by
\begin{equation}
\label{cl2}
Z_2(\zeta,\Lambda) \, m^{\rm bm}(\Lambda) \equiv 0\,,\; \forall \
\Lambda \gg \zeta\,,
\end{equation}
which is equivalent to requiring $\hat m = 0$ \cite{Maris:1997tm}, where $\hat m$ is the renormalization-point-invariant current-quark mass.  This means, of course, that we suppress the effect of electroweak interactions, which explicitly violate ${\rm SU}(2)_L \times {\rm SU}(2)_R$ chiral symmetry.

Equation~(\ref{fpigen}) is the exact expression in QCD for the pion's leptonic decay constant.\footnote{In the neighborhood of the chiral limit, a value for $f_\pi$ can be estimated via either of two approximation formulae \protect\cite{Pagels:1979hd,Cahill:1985mh,Chang:2009zb}.  These formulae both illustrate and emphasize the role of $f_\pi$ as an order parameter for DCSB.}
It is a property of the pion and, as consideration of the integral expression reveals, it can be described as the pseudo-vector projection of the pion's Bethe-Salpeter wavefunction onto the origin in configuration space.  (It can also be defined as the integral of the pion's gauge invariant distribution amplitude~\cite{Lepage:1980fj}.)  We note that the product $\psi = S \Gamma S$ is called the Bethe-Salpeter wavefunction because, when a nonrelativistic limit can validly be performed, the quantity $\psi$ at fixed time becomes the quantum mechanical wavefunction for the system under consideration.

If chiral symmetry were not dynamically broken, then in the neighborhood of the chiral limit $f_\pi \propto \hat m$ \cite{Holl:2004fr}.  Of course, chiral symmetry is dynamically broken in QCD \cite{Bhagwat:2003vw,Bhagwat:2006tu,Bowman:2005vx} and
\begin{equation}
\lim_{\hat m\to 0} f_\pi(\hat m) = f_\pi^0 \neq 0\,.
\end{equation}
Taken together, these last two observations express the fact that $f_\pi^0$, which is an intrinsic property of the pion, is a \emph{bona fide} order parameter for DCSB.
%PCT-----
A typical estimate from chiral perturbation theory \cite{Bijnens:2006zp} suggests that the chiral limit value, $f_\pi^0$, is $\sim 5$\% below the measured value of 92.4\,MeV; and efficacious DSE studies give a 3\% chiral-limit reduction~\cite{Maris:1997tm}.
%-------
In connection with the leptonic decay, it is interesting to note that $\Gamma_{\pi^+ \to \mu^+\nu} \propto f_\pi^2 m_\pi$.  In contrast, within a constituent-quark model, $\Gamma_{\pi^+ \to \mu^+\nu} \propto |\psi(0)|^2$, where $\psi(r)$ is the pion's constituent-quark wavefunction \cite{VanRoyen:1967nq}.  Therefore, consistency with DCSB in QCD requires that a realistic pion constituent-quark wavefunction must satisfy $|\psi(0)| \propto \sqrt m_\pi$ in the neighborhood of the chiral limit \cite{Nussinov:2008rm}.

Equation~(\ref{rhogen}) is kindred to Eq.\,(\ref{fpigen}); it is the expression in quantum field theory which describes the \emph{pseudoscalar} projection of the pion's Bethe-Salpeter wavefunction onto the origin in configuration space.  Thus it is truly just another type of pion decay constant. Moreover, the physics becomes transparent upon the consideration of its chiral-limit behavior.

Complementing the discussion in Ref.\,\cite{Casher:1974xd}, a rigorous definition of an ``in-hadron condensate" was presented in Refs.\,\cite{Maris:1997hd,Maris:1997tm}.  In our context it is given by
\begin{equation}
\label{inpiqbq}
-\langle \bar q q \rangle_\zeta^\pi \equiv -
f_\pi \langle 0 | \bar q \gamma_5 q |\pi \rangle
= f_\pi \rho_\pi(\zeta) =: \kappa_\pi(\hat m;\zeta)\,.
\end{equation}
Since the dressed-quark propagator has a spectral representation when considered as a function of $\hat m$ \cite{Langfeld:2003ye}, one can derive from the axial-vector Ward-Takahashi identity a collection of Goldberger-Treiman-like relations for the pion \cite{Maris:1997hd,Goldberger:1958vp}, the most important of which herein is
\begin{equation}
\label{bwti}
E_\pi(k;0) = \frac{1}{f_\pi^0} \,B_0(k^2)\,,
\end{equation}
where $B_0$ is the scalar part of the dressed-quark self-energy computed in the chiral-limit.
This is Eq.\,(10) of Ref.\,\protect\cite{Maris:1997hd}.  Equations~(10)--(13) therein state that if, and only if, chiral symmetry is dynamically broken, then the solution of the chiral limit one-body problem for the dressed-quark propagator completely determines the leading amplitude in the solution of the pion bound-state problem, and it tightly constrains the bound-state's subleading amplitudes.

Using Eq.\,(\ref{bwti}), one finds \cite{Maris:1997hd}
\begin{eqnarray}
\nonumber
\lim_{\hat m\to 0} \kappa_\pi(\hat m;\zeta)
& = &  Z_4(\zeta,\Lambda) \, {\rm tr}_{\rm CD}\int^\Lambda \!\!\!\! \mbox{\footnotesize $\displaystyle\frac{d^4 q}{(2\pi)^4}$} S_0(q;\zeta) \\
& = &  -\langle \bar q q \rangle_\zeta^0\,.
\label{qbqpiqbq0}
\end{eqnarray}
Thus the so-called vacuum quark condensate is, in fact, the chiral-limit value of the in-pion condensate; i.e., it describes a property of the chiral-limit pion.  One can therefore argue that this condensate is no more a property of the ``vacuum'' than the pion's chiral-limit leptonic decay constant.  Moreover, Ref.\,\cite{Langfeld:2003ye} establishes the equivalence of all three definitions of the vacuum quark condensate: a constant in the operator product expansion \cite{Lane:1974he,Politzer:1976tv}; via the Banks-Casher formula \cite{Banks:1979yr}; and the trace of the chiral-limit dressed-quark propagator.  Note that Eq.~(\ref{gmor}) is now readily obtained using Eqs.\,(\ref{gmorgen}), (\ref{inpiqbq}), (\ref{qbqpiqbq0}).

We reiterate that the in-pion condensate is essentially equivalent to the pion's leptonic decay constant.  As with $f_\pi(\hat m)$, $\kappa_\pi(\hat m;\zeta)$ has a spectral representation in terms of the current-quark mass and, for all values of that mass, it is a well-defined, gauge-invariant and properly renormalized function.
On one hand, the quantity $\kappa_\pi(\hat m;\zeta)$ has a non-zero value in the chiral limit if, and only if, chiral symmetry is dynamically broken; and the so-called vacuum quark condensate is merely the result obtained when evaluating $\kappa_\pi(\hat m;\zeta)$ at this single value of its argument.  It is not qualitatively different from $f_\pi(\hat m = 0)$.
On the other hand \cite{Ivanov:1998ms},
\begin{equation}
\label{kappainfinity}
\lim_{\hat m \to \infty} \kappa_\pi(\hat m;\zeta) = {\rm constant.}
\end{equation}

The result in Eq.\,(\ref{kappainfinity}) contrasts sharply with the behavior of the trace of the dressed-quark propagator, which is usually considered to provide the context for an extension of the vacuum quark condensate to $\hat m\neq 0$.  The latter quantity exhibits a quadratic divergence \cite{Langfeld:2003ye}; viz., $\int^\Lambda \!\! \mbox{\footnotesize $\frac{d^4 q}{(2\pi)^4}$} S_{\hat m}(q) \propto \hat m \Lambda^2$ for $\hat m \gg \Lambda_{\rm QCD}$.  As a function of $\hat m$, it is thus only rigorously defined at a single value of its argument; i.e., on a set of measure zero.  In QCD, therefore, $\int^\Lambda \!\! \mbox{\footnotesize $\frac{d^4 q}{(2\pi)^4}$}{\rm tr} S_{\hat m}$ alone provides no information about the current-quark-mass-dependence of the dynamical chiral symmetry breaking phenomenon.  That is better traced through other means \cite{Chang:2006bm}.

It is now clear that both $f_\pi(\hat m = 0)$ and $\kappa_\pi(\hat m=0;\zeta)$ are intrinsic properties of the bound-state, in this case an isovector pseudoscalar meson constituted from equally massive current-quarks.  They are also order parameters for DCSB and, as elucidated in Ref.\,\cite{Chang:2009at}, they characterize QCD's susceptibility to respond to the insertion of a pseudoscalar probe.  Their role as order parameters is conveyed through the dressed-quark mass function, $M(p^2) = B(p^2;\zeta)/A(p^2;\zeta)$, via the Goldberger-Treiman relations derived in Ref.\,\cite{Maris:1997hd} and exemplified in Eq.\,(\ref{bwti}).

The derivation of Eq.\,(\ref{qbqpiqbq0}) given here makes this connection manifest.  It also shows that it is the dynamical generation of a non-zero quark mass function in chiral-limit QCD which expresses DCSB most fundamentally.  The behavior of $M(p^2)$ is now well known.  For $\hat m = 0$ it is power-law suppressed at ultraviolet momenta; viz., $\sim 1/p^2$.  On the other hand \cite{Bhagwat:2003vw,Bhagwat:2006tu,Bowman:2005vx}, $M(0) \sim 0.3\,$GeV$=1/[0.66\,{\rm fm}]$.  Gluons also acquire a large dynamical mass \cite{Cornwall:1981zr}.  Indeed, even in quenched-QCD a momentum-dependent dynamical mass, $m_g(k^2)$, appears in the transverse part of the gluon 2-point function \cite{Bonnet:2001uh,Boucaud:2010gr}, with $m_g^{\rm Q}(0) \approx 0.35\,$GeV$=1/[0.55\,{\rm fm}]$.  The inclusion of dynamical quarks leads to an increased value \cite{Bowman:2004jm}; viz., $m_g(0)\sim 0.45\,$GeV, but with $m_g^{\rm Q}(k^2)=m_g(k^2)$ for $k^2\gtrsim 10\,$GeV$^2$.

Both of these dynamical mass functions owe their long-range enhancement to the non-Abelian nature of QCD and are intimately connected, as suggested by the similar magnitude of their values at infrared momenta; i.e.,
\begin{equation}
\label{M0mg0}
M(0) \sim m_G(0) \approx 0.4\,{\rm GeV} \approx 1/[0.5\,{\rm fm}] =: m_{\rm ir}\,.
\end{equation}
This mass-scale provides an infrared cutoff within an hadron, such that the role played by constituent field modes with $p^2 < m_{\rm ir}^2$ is suppressed.  This is the dynamically-induced infrared regulator we referred to in connection with Eqs.\,(\ref{fpigen}), (\ref{rhogen}).  Defining a ``wavelength'': $\lambda := 1/\sqrt{p^2}$, then an equivalent statement is that modes with $\lambda$ increasing beyond $\lambda_{\rm ir} = 1/m_{\rm ir}$ play a progressively smaller part in defining the bound-state's properties.
%The phenomenon of dynamical mass generation underlies the maximal wavelength discussed in Ref.\,\cite{Brodsky:2008be}.
The phenomenon of dynamical mass generation is closely related to the maximal wavelength condition discussed in Ref.\,\cite{Brodsky:2008be}; and both are consequences of confinement in QCD.
%(NB.\ For light-front longitudinal zero modes, this is a constraint on $|p_\perp|$.)
%
Recent DSE studies of ground-state pseudoscalar and vector quarkonia, from the $u$- to the $b$-quark region, and the corresponding $D$- and $B$-mesons, indicate that a DCSB-induced infrared regularization of both the self-interaction dressing and binding dynamics of $c$- and $b$-quarks tends to improve the computed values of hadron masses and electroweak decay constants \cite{Nguyen:2009if}.

As stated above, color confinement is important in establishing condensates as in-hadron rather than vacuum matrix elements.  This may be visualized via the $B$-meson; i.e., a bound-state of a heavy $\bar b$- and a light-quark.  Dyson-Schwinger equation and lattice-QCD studies demonstrate that the propagator for an apparently isolated light-quark acquires a momentum-dependent mass function \cite{Bhagwat:2003vw,Bhagwat:2006tu,Bowman:2005vx}, with which a non-zero condensate can be associated \cite{Langfeld:2003ye}.  However, in a fully self-consistent treatment of the bound state, this phenomenon occurs in the background field of the $\bar b$-quark, whose influence on light-quark propagation is primarily concentrated in the far infrared and whose presence ensures the manifestations of light-quark dressing are gauge invariant.
%%--notions of heavy-quark symmetry ... size of heavy-light hadron is fixed in limit m_Q->\infty

Equation~(\ref{qbqpiqbq0}) shows that the so-called vacuum quark condensate is the chiral-limit value of the in-hadron condensate in all reference frames.  It is a property of the bound-state in precisely the same manner as $f_\pi$.  It is therefore of interest to consider these two quantities in the light-front framework.  Following Ref.\,\cite{Lepage:1980fj}, one finds in the collinear frame; i.e., $P=(P^+,P^-=m_\pi^2/P^+,\vec{0}_\perp)$:
\begin{eqnarray}
\nonumber
f_\pi P^- & = & 2 \sqrt N_c \, Z_2 \int_0^1\! dx \! \int\mbox{\footnotesize $\displaystyle\frac{d^2 k_\perp}{16\pi^3}$} \, \psi(x,k_\perp) \frac{k_\perp^2\! + m_\zeta^2}{P^+\,x(1-x)}\\
&& + \mbox{instantaneous}\,,
\label{LFfpi} \\
\nonumber
\rho_\pi & = & \sqrt N_c \, Z_2 \int_0^1\! dx\!
\int\mbox{\footnotesize $\displaystyle\frac{d^2 k_\perp}{16\pi^3}$} \,
\psi(x,k_\perp) \, \frac{ m_\zeta }{x (1-x)}\\
&& + \mbox{instantaneous}\,,
\label{LFrhopi}
\end{eqnarray}
where both currents receive contributions from the ``instantaneous'' part of the quark propagator ($\sim \gamma^+/k^+$) and the associated gluon emission, which are not written explicitly.  In Eqs.\,(\ref{LFfpi}) and (\ref{LFrhopi}), $\psi(x,k_\perp)$ is the valence-only Fock state of the pion's light-front wavefunction.  Its integral over transverse momentum defines the gauge-invariant pion distribution amplitude.

Recall now that the pion's leptonic decay constant is an order parameter for DCSB.  In the context of Eq.\,(\ref{fpigen}), this is readily seen: owing to Eq.\,(\ref{bwti}) and linearity of the Bethe-Salpeter equation in the solution $\Gamma_\pi$, the magnitude of $f_\pi$ is determined by that of the scalar piece of the dressed-quark self-energy.  This is large in the chiral limit when chiral symmetry is dynamically broken, but otherwise vanishes for $\hat m =0$.
Furthermore, from the discussion associated with Eqs.\,(\ref{rhogen}), (\ref{inpiqbq}) and (\ref{qbqpiqbq0}), we have already seen that $\rho_\pi$ is also a DCSB order parameter.

However, how are $f_\pi(\hat m = 0)$ and $\rho_\pi(\hat m = 0)$ to be established as order parameters from Eqs.\,(\ref{LFfpi}) and (\ref{LFrhopi})?
% The answer should somehow be encoded in $\psi$.
%existence or absence of zero modes ... order parameter for DCSB.
This aspect of these in-hadron properties is difficult to see from these equations.  For example, the right-hand-side of Eq.\,(\ref{LFrhopi}) gives the appearance of vanishing identically in the chiral limit, which is defined in Eqs.\,(\ref{cl1}), (\ref{cl2}).

Within the light-front formulation of QCD, the question of DCSB has often led to a consideration of longitudinal zero modes.  That discussion has almost exclusively been conducted in the context of producing: (1) a non-zero value for the vacuum quark condensate; and (2) a Goldstone boson. (See, e.g., Refs.\,\cite{martinovic,Wu:2003vn,Ji:2005wd}.)
However, with a shift in the paradigm so that DCSB is understood as being expressed in properties of hadrons rather than of the vacuum, it becomes apparent that zero modes cannot provide for nonzero values of $f_\pi$ and $\rho_\pi$ in Eqs.\,(\ref{LFfpi}) and (\ref{LFrhopi}). Indeed, owing to confinement, light-front modes with $k^+ \ll m_{\rm ir}$ are exponentially suppressed within a bound-state and hence cannot contribute materially to these in-hadron properties.

\begin{figure}[t]
\includegraphics[clip,width=0.32\textwidth]{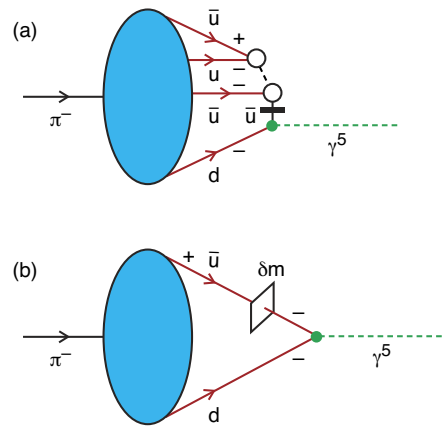}
\caption{\label{instantaneous} (Color online)
Light-front contributions to $\rho_\pi=-\langle 0| \bar q \gamma_5 q |\pi\rangle$.
\emph{Upper panel} -- A non-valence piece of the meson's light-front wavefunction, whose contribution to $\rho_\pi$ is mediated by the light-front instantaneous quark propagator (vertical crossed-line).  The ``$\pm$'' denote parton helicity.
\emph{Lower panel} -- There are infinitely many such diagrams, which can introduce chiral symmetry breaking in the light-front wavefunction in the absence of a current-quark mass.
(The case of $f_\pi$ is analogous.)}
\end{figure}

An alternative is illustrated in Fig.\,\ref{instantaneous}.  The light-front-instantaneous quark propagator can mediate a contribution from higher Fock state components to the matrix elements in Eqs.\,(\ref{LFfpi}) and (\ref{LFrhopi}).  Such diagrams connect dynamically-generated chiral-symmetry breaking components of the meson's light-front wavefunction to these matrix elements.  There are infinitely many contributions of this type and they do not depend sensitively on the current-quark mass in the neighborhood of the chiral limit.  Thus, DCSB in the light-front formulation, expressed via in-hadron condensates, is seen to be connected with sea-quarks derived from higher Fock states.  This solution is kindred to that discussed in Ref.\,\cite{Casher:1974xd}.  We note that the role of the instantaneous contributions is also emphasized in Ref.\,\cite{Wu:2003vn}, with the suggestion that they conspire to ensure Eqs.\,(\ref{LFfpi}) and (\ref{LFrhopi}) are consistent with Eq.\,(\ref{gmorgen}).

Herein we have presented an alternate view of the chiral order parameter which is conventionally understood as the vacuum quark condensate; namely, that it is qualitatively equivalent to the pion decay constant and is localized within the hadron.  There are also other QCD quantities that are conventionally interpreted as vacuum condensates, uniform in spacetime, such as $\langle G_{\mu\nu}G^{\mu\nu}\rangle$.  Condensates are typically introduced as \emph{a priori}-undetermined mass-dimensioned parameters in the operator product expansion of a color-singlet current-current correlator.  As such, the so-called vacuum condensates have come to be useful theoretical devices.  However, this should not be permitted to obscure the fact that their rigorous definition is delicate, owing, e.g., to possible dependence on the normal-ordering prescription; nor that current-current correlators can be calculated in a manner that does not involve the introduction of vacuum condensates, such as via Dyson-Schwinger equations \cite{Fischer:2006ub,Roberts:2007jh,Chang:2010jq} or lattice gauge theory \cite{Bazavov:2009bb}.

%\begin{acknowledgments}
%
This study was conceived at a workshop sponsored by the
Argonne/U.\,Chicago Joint Theory Institute, funded by ANL's
%Laboratory-Directed Research and Development
LDRD program.  We acknowledge valuable discussions with P.\,O.~Bowman during this event and subsequent conversations with Guy de T\'eramond.
This work was supported in part by:
U.\,S.\ Department of Energy contract no.~DE-AC02-76SF00515;
U.\,S.\ Department of Energy, Office of Nuclear Physics, contract no.~DE-AC02-06CH11357;
and the U.\,S.\ National Science Foundation, under grants
% (R.S.)
NSF-PHY-06-53342
% (P.C.T.).
and NSF-PHY-0903991.
SJB also thanks the Hans Christian Andersen Academy and Professor Franceso Saninno for hosting  his visit at CP$^3.$
%\end{acknowledgments}

%PCT-------
%\input BRSTbib

%-----------
\end{document}